\documentclass[aps, prl,twocolumn,
                showpacs,preprintnumbers,superscriptaddress]{revtex4}

\usepackage{epsfig}
\usepackage{graphicx}
\usepackage{dcolumn}
\usepackage{bm}

\begin{document}

\title{First-Principles Description of Correlation Effects in Layered 
Materials}

\author{Andrea Marini}
\affiliation{Istituto Nazionale per la Fisica della Materia e
Dipartimento di Fisica dell'Universit\'a di Roma ``Tor Vergata'',
Via della Ricerca Scientifica, I-00133 Roma, Italy}
\affiliation{European Theoretical Spectroscopy Facility (ETSF)}

\author{P. Garc\'{\i}a-Gonz\'alez}
\affiliation{European Theoretical Spectroscopy Facility (ETSF)}
\affiliation{Departamento de F\'{\i}sica Fundamental, 
Universidad Nacional de Educaci\'on a Distancia, Aptd. 60.141, E-28080 Madrid, Spain}

\author{Angel Rubio}
\affiliation{European Theoretical Spectroscopy Facility (ETSF)}
\affiliation{Institut f\"ur Theoretische Physik, Freie Universit\"at Berlin,
Arnimallee 14, D-14195 Berlin, Germany}
\affiliation{ 
Departamento de F\'{\i}sica de Materiales, 
Facultad de Qu\'\i micas Universidad del Pa\'\i s Vasco, 
Centro Mixto CSIC-UPV, and Donostia International Physics Center (DIPC), 
E-20018 Donostia-San Sebasti\'an, Spain}

\date{\today}

\begin{abstract}
We present a first-principles description of anisotropic materials 
characterized by having both weak (dispersion-like) and strong 
covalent bonds, based on the Adiabatic--Connection 
Fluctuation--Dissipation Theorem within Density Functional Theory.
For hexagonal boron nitride
the in-plane and out of plane bonding as well as vibrational dynamics are
well described both at equilibrium and when the layers are pulled 
apart. Also bonding in covalent and ionic solids is described. 
The formalism allows to ping-down the deficiencies of common
exchange-correlation functionals and provides insight towards  the 
inclusion of dispersion interactions into the correlation functional.
\end{abstract}


\maketitle

Soft matter, biomolecules, and layered materials are examples of weakly 
bonded compounds, which constitute stringent tests for \emph{ab-initio} 
electronic structure calculations. 
A successful description of such systems requires an accurate treatment of 
a whole range of interactions, from short-range covalent to long-range
van der Waals\,(vdW) forces. 
Moreover, their study provides valuable hints about how to devise new 
approximations that describe at a reasonable computational 
cost very different bonding regimes.
In this realm, layered materials represent a unique case. The 
Kohn-Sham\,(KS) method under the simple local density 
approximation\,(LDA)~\cite{KS65} is known to give reasonable structural 
properties \emph{near the equilibrium interlayer distance} of systems like 
graphite~\cite{CGM94} and hexagonal boron 
nitride\,(\emph{h}-BN)~\cite{KKH99,YLC03}.
Surprisingly the more elaborated 
exchange-correlation (XC) energy functional based on the
generalized-gradient approximation\,(GGA)~\cite{PBE96}, 
while provides a good in-plane bonding
fails dramatically in the description of
the equilibrium interlayer distance in \emph{h}-BN and graphite, 
leading to completely unphysical results~\cite{DLR01,RDJ03}.
The mechanism responsible for this poor performance 
is not specific to GGA but to all functionals based on
exact--exchange~\cite{exx1p} plus some LDA correlation\,(EXXc) (as 
illustrated below).

It has been hypothesized that this failure is due to the lack 
of description of vdW forces, which are the manifestation of long-range
correlation effects~\cite{RDJ03}. However LDA, without accounting for
vdW interactions, does bind the \emph{h}-BN and graphite layers.
More importantly, recently proposed approximate non--local prescriptions 
beyond LDA or GGA in the vdW Density Functional\,(vdW--DF) framework~\cite{RLLD00}, 
while describing vdW forces, result in a 
large overestimation of the equilibrium layer--layer distance~\cite{RDJ03}. 
Thus, available XC functionals cannot describe 
correctly a layered material at different geometrical configurations. 
Consequently, there is a basic need to 
understand this severe breakdown and how to define a stable, 
successful scheme to calculate such structural properties.
This is the goal of the present Letter. We demonstrate that a fully 
microscopic, parameter--free approach based on the Adiabatic--Connection 
Fluctuation--Dissipation Theorem\,(ACFDT)~\cite{LP75} correctly describes
the mechanical properties of \emph{h}-BN \emph{near and far from} the 
equilibrium configuration. We show that ACDFT does bind the \emph{h}-BN layers
at an equilibrium distance in excellent agreement with the
experiment, accounting  by construction for vdW forces~\cite{HAL96,DW99}.
In contrast, while LDA overbinds \emph{h}-BN, GGA, EXXc and, obviously,
any other prescription with  a local or semi-local correlation 
part fail in describing the layers bonding.

\emph{h}-BN has a wealth of important applications for high resistance and 
inert nanodevices~\cite{KKH99,YLC03,Rap85}. 
Furthermore, the \emph{h}-BN layer-layer interaction is weaker than in 
graphite (the paradigmatic example of layered material), thus being a 
better scenario to assess and understand the performance
of KS-based schemes. The ACFDT has been applied so far only
to jellium systems~\cite{DW99}, molecules~\cite{FG02},
or compact solids~\cite{MAK02}. However it starts to be
consider a promising \emph{ab-initio} method because:
i) it provides a reliable
first-principles description of XC effects at a lower numerical cost than
Quantum Chemistry methods or statistical Quantum Monte-Carlo techniques;
ii) it is naturally built into the KS scheme 
as a way to construct \emph{advanced} XC energy functionals
depending on the electron density $n\left( \mathbf{r}\right)$ \emph{and} 
on the set of occupied [$\phi_{v}\left( \mathbf{r}\right)$] and empty 
[$\phi_{c}\left( \mathbf{r}\right)$] KS orbitals. Indeed, according to the 
ACFDT, the total XC energy comprises of the exact-exchange energy
$E_{\mathrm{X}} =-\frac{1}{2} \int d\mathbf{r} d\mathbf{r'}
\left| \sum_{v}^\mathrm{occ} \phi _{v}^{\ast }\left( 
\mathbf{r}\right) \phi _{v}\left( \mathbf{r'}\right) \right|
^{2}/\left| \mathbf{r}-\mathbf{r'}\right|,
\label{4}
$
plus the correlation energy given \emph{exactly} by
\begin{equation}
E_{\mathrm{C}} =-\int_{0}^{+\infty }\frac{du}{2\pi }
\int_{0}^{\lambda }d\lambda \,\mathrm{Tr}\left( \widehat{w}\left[ 
\widehat{\chi }_{\lambda }\left( iu\right) -\widehat{\chi }_{0}
\left(iu\right) \right] \right),
\label{1}
\end{equation}
(we will use atomic units unless otherwise stated).
Tr denotes the spacial trace, and 
$\widehat{\chi}_{\lambda}\left(iu\right)$ is the density-density response 
function at imaginary frequencies of a fictitious electron-system 
interacting through a scaled Coulomb potential $\lambda w\left( 
\mathbf{r}\right) =\lambda /r$ whose ground-state
density equals the actual one. Thus, $\widehat{\chi}_{0}\left(iu\right)$
is the non-interacting response function of this fictitious KS system:
\begin{equation}
\chi_{0}\left(\mathbf{r},\mathbf{r'};iu\right)= 2 \Re \sum_{v,c}
\frac{
\phi_{v}^{\ast}\left(\mathbf{r}\right) \phi_{v}\left(\mathbf{r'}\right)
\phi_{c}^{\ast}\left(\mathbf{r'}\right) \phi_{c}\left(\mathbf{r}\right)
}
{iu+\left( \varepsilon_{v}-\varepsilon_{c}\right)},
\label{2}
\end{equation}
where $\varepsilon_{v}$ ($\varepsilon_{c}$) are the occupied (empty)
KS eigenenergies.  The interacting response $\widehat{\chi}$
can be obtained in the framework of time-dependent Density Functional 
Theory\,(TDDFT)~\cite{RG84,ORR02} by solving a Dyson-like equation:
\begin{equation}
\widehat{\chi }_{\lambda}\left( iu\right) =\left( 
\widehat{1}-\widehat{\chi }_{0}\left( iu\right) 
\left[ \lambda \widehat{w}+\widehat{f}_{\mathrm{XC},\lambda }
\left( iu\right) \right] \right) \widehat{\chi }_{\lambda}\left( iu\right),
\label{3}
\end{equation}
where $\widehat{f}_{\mathrm{XC},\lambda }\left( iu\right) $ is the XC kernel
of the fictitious system with the scaled interaction $\lambda \widehat{w}$. 
The Random Phase Approximation\,(RPA) to $\widehat{\chi }$
is obtained by setting $\widehat{f}_{\mathrm{XC}}=0$. 
Note that, in contrast to other procedures
put forward in the literature~\cite{RDJ03,exx1p}, this scheme provides a 
correlation functional built at the same level than exact exchange.

The XC effects introduced by 
$\widehat{f}_{\mathrm{XC}}$ on $\widehat{\chi }$ for the the calculation
of $E_{\mathrm{C}}$ are considerably weaker than for the description of the
optical properties of \emph{h}-BN~\cite{bn-opt},
as $\widehat{\chi}(iu)$ is a very smooth function
on the imaginary axis~\cite{footnote_Fxc}.
Therefore, we set 
$\widehat{f}_{\mathrm{XC},\lambda }\left( iu\right) =0$ in Eq.\,(\ref{3})
and correct the well known deficiency of the
RPA total correlation energy~\cite{LGP00} 
via a LDA-like correction term (a prescription often labeled
as RPA+) \cite{KP99}.
This scheme will be denoted in what follows as EXX/RPA+.  
Moreover, instead of solving selfconsistently the
KS equations for the EXX/RPA+ potential~\cite{NFG03}, we
evaluated the XC energy using the KS-LDA electronic structure 
as commonly done in Ab--initio Many--Body calculations~\cite{ORR02}.
We have explicitly checked that the use of GGA or EXXc KS states as
a basis for the EXX/RPA+ expressions has little effect (less than 2\,$\%$)
on the final XC energy.

Some technical details are relevant. The KS response function
is computed in reciprocal space using well converged KS orbitals and
eigenenergies obtained from the \textsc{abinit} code~\cite{abinit}. 
The interacting responses $\widehat{\chi}_{\lambda}$ are then obtained 
for each vector in the Brillouin zone (BZ), imaginary frequency $u$, and 
coupling parameter $\lambda $ using the \textsc{self} code~\cite{self}. 
Finally, the RPA correlation energy is calculated 
according to Eq.\,(\ref{1}) using a sixth order 
Gauss-Legendre (GL) sampling for $\lambda$ integration. The 
imaginary frequency integrals are done using two GL grids, each typically 
comprising 24 points. This allows a well converged evaluation of the 
contributions from both small and large imaginary frequencies. The 
convergence with respect to the BZ sampling has been also
carefully checked (the exchange energy is very sensitive to this
sampling). 
Nonetheless, the critical convergence parameters from a 
computational point of view are the number of
unoccupied bands and reciprocal space vectors in Eq.\,(\ref{2}).
The use of a plane-wave representation allows a systematic 
\emph{simultaneous} convergence with respect to these two parameters.
Further technical details, together with more applications to
metallic systems and graphite, will be presented elsewhere.

\begin{figure}
\begin{center}
\epsfig{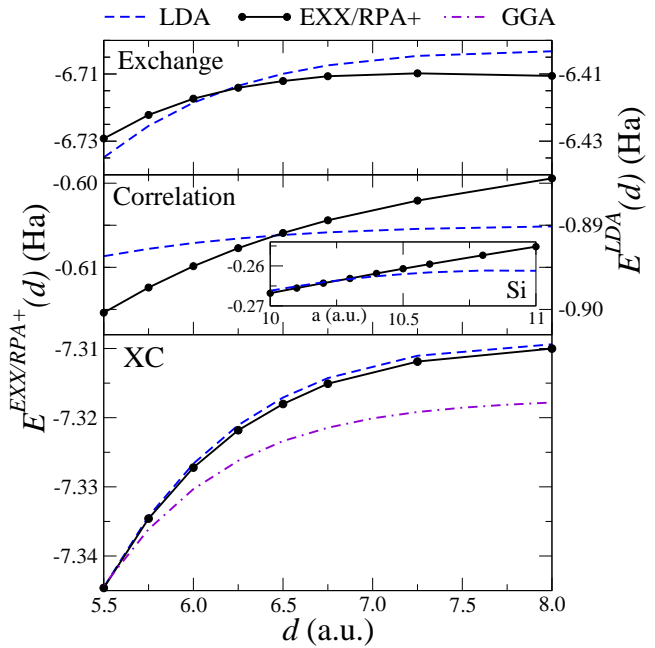}
\end{center}
\caption{(Color on--line) The LDA, EXX/RPA+, and GGA
exchange and correlation energies per unit cell for \emph{h}-BN
as a function of the layer-layer distance $d$
(the correlation energy of $Si$ as a function of the lattice constant $a$
 is showed in the inset).}
\label{Fig1}
\end{figure}

Before studying the weak interlayer interaction in \emph{h}-BN,
it is illustrative to analyze the impact of the exchange-correlation
effects at the EXX/RPA+ level for two prototype compact structures:
bulk Si (as covalent semiconductor) and NaCl (as ionic material).
It is known that the standard XC functionals, like GGA, 
provide a good description of those systems.
This is confirmed in the present ACFDT scheme as the impact in the
lattice constant of the EXX/RPA+ XC energy correction is below 0.5\,$\%$.
However, the hybrid EXXc overcorrects the GGA results 
specially in the low density regime of bulk Si due to the LDA
overestimation of the correlation energy for low densities 
(see inset of Fig.\ref{Fig1}). Nevertheless, this does not 
compromise very much the accuracy of the EXXc 
approach since in this case the exchange energy dominates
($E_\mathrm{X}\left(11\right)-E_\mathrm{X} \left(9.6\right)\approx0.2 \, 
\mathrm{Ha}\gg 
E_\mathrm{C}\left(11\right)-E_\mathrm{C}\left(9.6\right)\approx0.01 \, 
\mathrm{Ha}$)~\cite{note1}.

\begin{figure}
\begin{center}
\epsfig{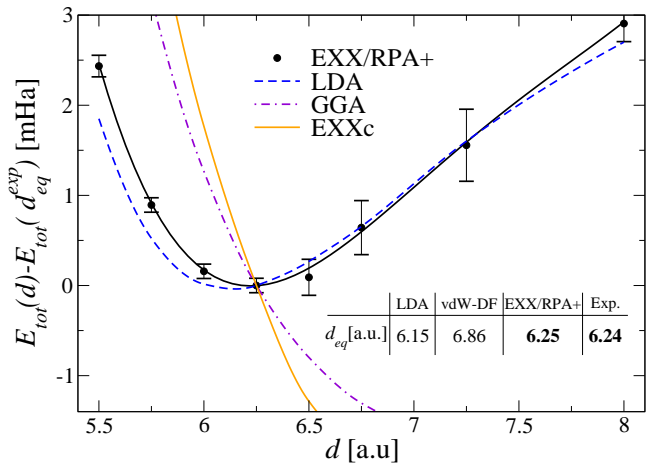}
\end{center}
\caption{(Color on--line) Variation of the total energy per unit cell of \emph{h}-BN versus
the layer-layer distance $d$. While EXX/RPA+ binds the \emph{h}-BN  layers, 
GGA and EXXc dramatically fail. EXX/PRA+ improves the equilibrium interlayer distance
compared to vdW--DF~\cite{RDJ03} and LDA, whose good performance is merely due to the 
cancellation of errors. The EXX/RPA+ error bars come from the extrapolation 
technique used to extract the correlation energy from values corresponding to finite number 
of bands and reciprocal space calculations. The experimental equilibrium distance
is taken from Ref.~\cite{bnexp}}
\label{Fig2}
\end{figure}

The previous scenario, where the variations of the exchange energy is at 
least one order of magnitude larger than correlation, changes completely 
in the case of \emph{h}-BN.
As displayed in Fig.\,\ref{Fig1}, the variations of 
$E_\mathrm{X}$ and $E_\mathrm{C}$ around the equilibrium interlayer 
distance $d_{eq}$ in \emph{h}-BN are of the same order of magnitude (few mHa) 
and at least one order of magnitude smaller than for Si or NaCl. Thus, 
small errors are more clearly visible and correlation plays a much more 
important role than in the case of the compact 
structures. Hence, a consistent description of exchange and
correlation is needed  to reproduce the delicate energetic 
interplay in the weak interlayer bonding of \emph{h}-BN.
This consistency is absent in the EXXc prescription and, as a
consequence, it provides too large (or even unbound) interlayer
equilibrium distances, as showed in Fig.\,\ref{Fig2}.
To a lesser extent, this problem is also present in the vdW--DF
scheme designed to treat vdW forces in an approximate manner\,\cite{RDJ03} that
largely overestimates the equilibrium distance ($6.86$ compared 
to $6.24$\,a.u.), while underestimating the bulk modulus.
The GGA follows the very same trend than the EXXc. 
This puts in evidence the fact that GGA exchange greatly 
improves upon the LDA (and, hence, provides better chemical properties 
of molecules), but does not do a so good job for correlation. 
Instead EXX/RPA+, which treats consistently exchange \emph{and} correlation,
correctly predicts \emph{h}-BN to be stable, with structural parameters 
in excellent agreement with experiment (see Fig.\,\ref{Fig2}).
Furthermore, in contrast to LDA, GGA, and EXXc, EXX/RPA+ naturally
embodies the correct long-range $d^{-4}$ vdW power law dependence of the
energy~\cite{DLR01} as a function of interlayer distance,
thus merging continuously the covalent and vdW regimes. 
This is confirmed in Fig.\,\ref{Fig2} where,
for $d$ larger than $7.5$\,a.u., we observe that LDA deviates from 
the correct total energy dependence  (given by EXX/RPA+).
By looking at the LDA XC contributions in Fig. \ref{Fig1},
we see that even if 
both $E_\mathrm{X}$ and $E_\mathrm{C}$ are poorly described, the 
XC energy dependence on the interlayer distance  is
similar to the ACDFT result.
Nevertheless, the slope of the LDA XC energy is larger than the 
ACDFT one. This difference is
responsible for the smaller interplane distance in the LDA as compared
to EXX/RPA+, being the latter ($6.25$\,a.u.) in stringent agreement with the 
experimental value $d_{\mathrm{eq}}^{\mathrm{exp}} = 6.24$\,a.u.~\cite{bnexp}.
Moreover in the ACFDT the LDA bulk modulus B is increased by about 5\% 
in better agreement with the experimental value
(in contrast to the vdW-DF scheme~\cite{RDJ03} where B is reduced).
Therefore, while the relative 
success of the LDA around the equilibrium configuration is merely due to a 
cancellation of errors the EXX/RPA+ scheme provides better
structural properties describing equally well the correlation and the
exchange energy. This shows that the ACFDT Ab--Initio scheme  is
suitable to study very different regimes, like Si and NaCl, where the
exchange energy dominates, or like \emph{h}--BN, where the success
is bound to a correct balance of correlation and exchange contributions.

An additional severe test of the accuracy of the EXX/RPA+ method
is given by the calculation of  the vibrational spectra of \emph{h}-BN. 
LDA and GGA in--plane lattice dynamics~\cite{KKH99} reproduces
fairly well the available experimental results. Similarly, 
the results of a frozen--phonon calculation 
within the present EXX/RPA+ scheme give in--plane modes at the high-symmetry points within
few percents from experiments (and LDA or GGA).
The situation changes considerably for interplane modes where 
the LDA and GGA phonons differ
by up to 30\,$\%$ when calculated at the experimental interplane distance
(this is the only way to compare the results as GGA does not bind
\emph{h}-BN and LDA overbinds).
Indeed, modes related
to interlayer vibrations would be very much sensitive to errors
in the correlation functional (as the typical variations in the correlation
energy that has to be resolved are up to two orders of magnitude
smaller for interlayer phonons than for structural characterization).
Furthermore, recent second-order Raman experiments~\cite{RFA05}
have suggested that the interplane 
modes are greatly underestimated in the theoretical calculations.
In particular they quote a 
frequency value for the $\Gamma_{4}^{+}$ phonon mode (where two BN planes 
rigidly slide against each other) of 310 cm$^{-1}$, which is about a
factor of three larger than the LDA (90 cm$^{-1}$) or GGA (122 cm$^{-1}$)
values\cite{note-phonon}, computed at the experimental lattice parameters.
If the experimental analysis based on overtones was correct it 
would imply that correlation effects (dispersion forces)
are very important for those vibrations. This point can be directly
addressed by the present EXX/RPA+ scheme as we have shown that it describes
properly the structural properties of \emph{h}-BN and accounts at the same time  by
construction for the long-range vdW forces.
Surprisingly, the calculated EXX/RPA+ $\Gamma_{4}^{+}$ phonon
mode energy  is $130\pm10$ cm$^{-1}$, regardless the use of the theoretical or
experimental lattice parameters. Therefore, it is clear that
improving the description of XC effects upon LDA brings the $\Gamma_{4}^{+}$
stiffer, still not close to the experimental assignment.
However, this huge discrepancy can be drastically reduced if in the
analysis of the experimental data 
we allow for the whole set of double--phonon processes (sum and difference)
consistent with the selection rules of Raman scattering
instead of considering only phonon-overtones.
If we do so, the new frequency assignment for the $\Gamma_{4}^{+}$ phonon mode
is  about 125 cm$^{-1}$~\cite{2ph_note},
still in disagreement with LDA value but very close to the EXX/RPA+ result
and to the GGA value (in spite of the fact that it
does not bind \emph{h}-BN).
Therefore, EXX/RPA+ solves unambiguously this puzzling situation providing
an accurate method to evaluate the role played by the exchange--correlation
effects.

In summary, we have shown that EXX/RPA+ correctly describes the
structural properties, the on--plane and  the out--of--plane phonon modes 
of \emph{h}-BN.
While both GGA and EXXc do not predict the \emph{h}-BN layers
bonding and LDA underestimates both the interlayer phonon mode 
frequency and interlayer distance,
EXX/RPA+ provides a consistent, accurate and successful alternative scheme.
Thus, we have seed light about the performance of standard \emph{ab-initio}
techniques when applied to layered systems where
any improvements upon LDA require equally accurate treatments of
exchange and correlation effects.
The EXX/RPA+ scheme presented here opens the field to
more elaborated \emph{ab-initio} applications in the wide field 
of  sparse systems and soft-matter.

The authors were partially supported by the EC
Network of Excellence NANOQUANTA (NMP4-CT-2004-500198) and Spanish MCyT 
(FIS2004-05035-C03-03).
PGG thanks the Ramon y Cajal Program and AR the 
2005 Bessel research award of the Humboldt Foundation.
We benefited from discussions with J.F. Dobson, J. Jung, R.W.
Godby, M. Gr\"uning, J. Serrano, and L. Wirtz.


\begin{thebibliography}{99}
\bibitem{footnote_Fxc}
 For structural
 properties (i.e. energy differences in isoelectronic systems), there are
 evidences indicating that $\widehat{f}_{\mathrm{XC}}$ plays
 a minor role~\cite{DW99,FG02,KP99}.
\bibitem{note1} 
 As it is well known, 
 LDA underestimates (overestimates) the absolute value of the exchange 
 (correlation) energy. Both errors compensate and the LDA
 and ACFDT XC energies are very similar for compact structures.
\bibitem{note-phonon}
 We should stress that if the LDA phonon calculation
 is performed at the theoretical lattice constant,
 the corresponding $\Gamma_{4}^{+}$ phonon frequency is 120 cm$^{-1}$,
 in rather good agreement with the new assignment of the 
 experimental $\Gamma_{4}^{+}$ phonon mode \cite{2ph_note}.
\end{thebibliography}
\end{document}